\documentclass[showpacs,floatfix,amsmath,superscriptaddress,amssymb,prb,twocolumn]{revtex4}
\usepackage{latexsym}
\usepackage{graphicx}

\newcommand{\be}{\begin{equation}}
\newcommand{\ee}{\end{equation}}
\newcommand{\ba}{\begin{eqnarray}}
\newcommand{\ea}{\end{eqnarray}}
\newcommand{\baa}{\begin{eqnarray*}}
\newcommand{\eaa}{\end{eqnarray*}}

\begin{document}
\title{The effects of non-abelian statistics on two-terminal shot noise in
a quantum Hall liquid in the Pfaffian state}
\author{Cristina Bena}
\affiliation{Department of Physics and Astronomy, University of
California, Los Angeles, California 90095-1547}
\author{Chetan Nayak}
\affiliation{Department of Physics and Astronomy, University of
California, Los Angeles, California 90095-1547}
\affiliation{Microsoft Research, Project Q, Kohn Hall, University of California,
Santa Barbara, CA 93108}

\date{\today}

\begin{abstract}
We study non-equilibrium noise in the tunnelling current between
the edges of a quantum Hall liquid in the Pfaffian state, which is a strong candidate for
the plateau at $\nu=5/2$. To first non-vanishing order in perturbation theory (in the tunneling amplitude) we find that one can extract the value of the fractional charge of the tunnelling quasiparticles. We note however that no direct information about non-abelian statistics can be retrieved at this level. If we go to higher-order in the
perturbative calculation of the non-equilibrium shot noise,
we find effects due to non-Abelian statistics.
They are subtle, but eventually may have an experimental signature on the frequency dependent shot noise.
We suggest how multi-terminal noise measurements might yield a more dramatic
signature of non-Abelian statistics and develop some of the relevant formalism.
\end{abstract}

\pacs{73.23.-b, 71.10.Pm, 73.43.Jn }

\maketitle

\section{Introduction}
Recently, there has been increased interest in the possibility that
non-Abelian braiding statistics might occur in the fractional quantum
Hall regime, particularly at  $\nu =5/2$ and, possibly, other fractions
in the first excited Landau level. One possible state describing
the $\nu=5/2$ plateau is the Pfaffian state \cite{pfaffian1,pfaffian2,pfaffian3,pfaffian-eft}, which combines
aspects of the fractional quantum Hall effect and  BCS pairing. Its quasiparticle
excitations obey non-Abelian braiding statistics \cite{pfaffian1,pfaffian2,pfaffian3,pfaffian-eft}.
Thus far, the strongest evidence in favor of this state comes from numerical
studies of electrons in the first excited Landau level \cite{numerics}.
Given the special topological properties of this state, it would be of great importance to have a clean experimental test to identify it. Such tests, measuring directly the braiding of quasiparticles,
have been proposed in refs. \onlinecite{pfaffian-eft,DasSarma05,Stern05,Bonderson05}.
Here, we see if more indirect, but simpler experiments can reveal some of the same properties of this state.

Current noise measurements are an interesting possibility. Since the
current is delivered in fractional charge $e^*$ packets, the zero-frequency
limit of the current noise evinces shot noise. The ratio between the shot
noise and the current is simply $e^*$, as has been observed
experimentally \cite{shotexp}.
It is believed \cite{shotproposals,eunah} that such experiments
could also be used in various setups to also "measure" the statistics of
quasiparticles, though actual experiments addressing this issue are much fewer\cite{statexp}.
In photon-counting experiments, ``bunching'' is observed, while the analogous
noise measurments in a Fermi liquid show ``anti-bunching'', as a result of
Fermi statistics. In a state in which the quasiparticles are anyons, one might expect
something intermediate between these two limits, from which the statistics
might be extracted.
In the Pfaffian state, something even more interesting might be observed
since the quasiparticles have non-Abelian statistics.

The difference between a non-Abelian state, such as the Pfaffian, and an
Abelian state, such as the Laughlin state, is most easily seen when
two quasiparticles are brought together, or `fused'. In an Abelian state,
there is a unique quasiparticle type which is obtained (up to additional bosonic
excitations). For instance, when two $e/3$ Laughlin quasiparticles are
brought together, a charge $2e/3$ quasiparticle is obtained. When a
charge $e/3$ quasiparticle is taken around a $2e/3$ quasiparticle, the wavefunction
acquires a factor $e^{4\pi i/3}$. One the other hand, when two charge $e/4$ quasiparticles
in the Pfaffian state are fused together, there are {\it two possible outcomes}, both
with charge $e/2$. They differ by the presence or absence of a neutral fermion.
Thus, when a charge $e/4$ quasiparticle is taken around two fused
$e/4$ quasiparticles, the factor which is acquired is either $i$ or $-i$,
depending on how the particles fuse. (When they fuse to form a linear
combination of the two possible outcomes, taking an $e/4$ quasiparticle around
them would transform them into a different linear combination.)
It is this multiplicity of outcomes of fusion which
we would like to access through consideration of the noise.

In this paper we analyze the non-equilibrium noise in the
tunnelling current between two edges of a FQH liquid in a Pfaffian state and we
find that, indeed, information about the fractional charge and statistics of the FQH
Pfaffian quasiparticles can be extracted from the two-terminal noise.
In particular, we find that up to first order in perturbation theory, the shot noise contains sufficient information to allow for the measurement of the fractional charge of the tunnelling quasiparticles.
The charge also shows up in the Josephson frequency. When we also compute higher-order corrections to the noise, non-Abelian effects come into play. One of the effects of statistics in  the
Pfaffian state is that the peak in the noise at the Josephson frequency is shifted toward lower frequencies and enhanced, while in the Laughlin state it is flattened.

This is less dramatic than the underlying physics because once
the quasiparticle four-point function correlation function is Fourier transformed and
inserted into the expression for the noise, its structure is masked.
Nevertheless, a two-terminal noise measurement is a good starting
point for discussing more complicated setups.
We discuss other possible experiments, with more terminals, and
assess their usefulness for observing the effects of non-Abelian statistics.

In Section \ref{formalism} we lay out the formalism necessary to calculate the tunnelling current and current fluctuations and note the expected non-linear $I-V$ characteristic for tunneling
into the Pfaffian state.
In Section III we analyze the first non-zero order in perturbation theory for both the tunnelling current and shot noise. In Section IV we go to the next order in perturbation theory and to finite temperature. In Section V we present and discuss our results. We conclude in Section VI.

\section{Formalism}
\label{formalism}

The Pfaffian state \cite{pfaffian1,pfaffian2,pfaffian3,pfaffian-eft} describes electrons in
a half-filled Landau level (with straightforward generalization to other
even-denominator filling fractions or odd-denominator filling-fractions
for systems of bosons). While there does not appear to be a FQH plateau
at $\nu=1/2$, there is one at $\nu =5/2=2+\frac{1}{2}$. We ignore the 
filled lowest Landau level (of both spins), and focus on the half-filled
first excited Landau level, which we suppose is described by the Pfaffian state.
The gapless chiral theory describing excitations at the edge of a Pfaffian state is
a  $1+1-D$ conformal field theory (CFT) \cite{pfaffian-edge}.
The edge theory has a charge sector -- which is a free boson -- and a neutral sector
which is a $c=1/2$ Majorana fermion. The latter is the chiral part of the
critical theory of the $2D$ Ising model. In addition to the identity operator,
the latter has a spin or twist field $\sigma$ and a Majorana fermion operator $\psi$.

Considering the fact that a physical operator must
have a single-valued correlation function with the electron
operator, the operators which can be identified with
quasiparticle operators are $\Phi_{1/4}(x)= \sigma(x)\,e^{i \sqrt{g}
\phi(x)/2}$;  $\Phi_{1/2}(x)=e^{i \sqrt{g} \phi(x)}$;
$\Phi_{1}(x)=e^{i  \frac{1}{\sqrt{g}} \phi(x)}$; the neutral fermion $\psi$; 
and ${\Psi_e}(x) = \psi\,e^{i  \frac{1}{\sqrt{g}} \phi(x)}$. Here,
$g=1/2$.  Here
$\Psi_e$ is the electron annihilation operator, while $\Phi_{1/4}$, and
$\Phi_{1/2}$ annihilate quasiparticles with fractional charge $g e/2 =e/4$, and $g e= e/2$ respectively. In a tunnelling process between two edges of a quantum Hall liquid, the most relevant (in the RG sense) process will be the backscattering of the $e/4$ charged $\Phi_{1/4}$ quasiparticles. We need to note also that the fields $\phi$, $\psi$, and $\sigma$ also need to be characterized by a chiral index $R/L$, denoting which of the right/left moving branches of the theory the fields
belong to, such that  $\Phi_{1/4,R/L}(x)=e^{\pm i \sqrt{g}
\phi_{R/L}(x)/2} \sigma_{R/L}(x)$, and $\Psi^e_{R/L}(x)=\psi_{R/L} e^{\pm i  \frac{1}{\sqrt{g}} \phi_{R/L}(x)}$.

We can write an edge theory for the charged $\phi$ sector of the
theory, which is analogous to the Laughlin state with $g=1/2$. The
corresponding Lagrangian density for the right and left moving
modes is \cite{chamon,chamon2}:
\be
{\cal L}^{c}_{R/L}=\frac{1}{4 \pi} \partial_x
\phi_{R/L}(\pm
\partial_t+v\partial_x)\phi_{R/L},
\ee
where $v$ is the velocity of the edge excitations, which we will
set to $1$.
The fields $\phi_{R/L}$ satisfy the equal time commutation relations
\be
[\phi_{R/L}(t,x), \phi_{R/L}(t,y)]=\pm i \pi sgn(x-y).
\ee
We can also write \cite{chamon,chamon2} the total Lagrangian
in terms of $\phi=\phi_R+\phi_L$,
\be
{\cal L}_0(x,t) =\frac{1}{8 \pi} \{[\partial_t \phi(x,t)]^2-[\partial_x \phi(x,t)]^2\}.
\ee
where $\phi$ is satisfying $[\phi(t,x),\partial_t \phi(t,y)]=4 \pi i \delta(x-y)$.

The Lagrangian for the neutral sector of the theory is
\be
{\cal L}^{n}_{R/L}=\frac{1}{2 \pi}\, \psi_{R/L}(\pm
\partial_t+v\partial_x)\psi_{R/L},
\ee
The Ising spin fields $\sigma_{R/L}$ are twist fields for the Majorana fermion $\psi_{R/L}$:
whenever a Majorana fermion goes around its twist field, it changes sign.
The Ising spin  are most easily described in a conformal field theory picture.
The chiral $\sigma_{R/L}$ fields have conformal weight $1/16$, while the combination
$\sigma=\sigma_R \sigma_L$, which we will use later on, has dimension $1/8$.
Also, information about the two point correlations and four point correlation functions of the $\sigma$ fields have been derived in the CFT context. While the two-point functions
are simple power laws, the four point functions have a more complicated 
structure, which incorporates the effects of non-Abelian statistics.
This is described in detail in Appendix B.

\begin{figure}[h]
\begin{center}
\includegraphics[width=2.5in]{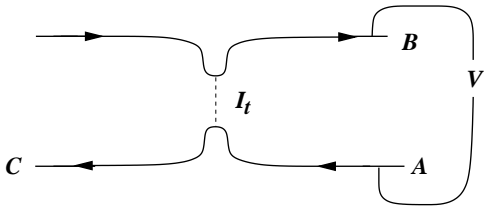}
\end{center}
\caption{The two-terminal noise setup in a quantum Hall bar. Current is
injected at $A$ and measured at $B$. The voltage drop between these two
terminals is also measured.}
\label{fig0}
\end{figure}

We are going to study a FQH setup similar to the one depicted in Fig.\ref{fig0}. In the presence of an applied voltage $V$, a current $I_0$ is injected in the sample through lead A.
In the absence of inter-edge tunnelling, this current will be picked up in lead C. If a gate voltage is applied across the sample, such that the sample is constricted, as indicated in the figure, quasiparticles from one edge can tunnel to the other edge, thereby giving rise to a tunnelling current $I_t$. This current can be measured in lead B. While the current in lead A will not be changed, the current picked up in lead C will be reduced to $I_0-I_t$.
 
We can write down the tunneling operator between the two edges of the quantum Hall fluid noting that a tunnelling process annihilates a right/left mover quasiparticle and creates a left/right moving quasiparticle on the opposite moving branch (see Fig.\ref{fig0}).
\vspace{-.1in}
\be
\label{eqn:tunnel-op}
H_{\rm tun}=\Gamma \Phi_{1/4,L}^{\dagger}(x=0) \Phi_{1/4,R}(x=0)+h.c.
\ee
which generates a change in the Lagrangian density:
\be
\delta {\cal L}(t,x)=-\Gamma \sigma^{\dagger}_R(t,0)
\sigma_L(t,0) e^{i \phi(t,0) \sqrt{g}/2}+h.c.
\label{eqn:tunnel-op-Lagrangian}
\ee
From the scaling dimensions of the fields in (\ref{eqn:tunnel-op}), we deduce that
the tunneling operator has dimension $3/4$. Hence, we expect the non-linear I-V
characteristic for weak inter-edge tunneling at $T=0$ to be \cite{pfcurrent}
\be
{I_t} \sim V^{-1/2} .
\ee
For $T>V$, the tunneling conductance varies as
\be
{G_T}\sim T^{-3/2} .
\ee

Note that the operator which tunnels a charge-1/2 quasiparticle is also relevant,
but less so than (\ref{eqn:tunnel-op}).
\begin{eqnarray}
\label{eqn:tunnel-op-subleading}
H_{\rm tun} &=& \Gamma' \Phi_{1/2,L}^{\dagger}(x=0) \Phi_{1/2,R}(x=0)+h.c.\cr
&=& -\Gamma e^{i \phi(t,0) \sqrt{g}}+h.c.
\end{eqnarray}

A voltage drop between the edges of the quantum Hall liquid can be introduced \cite{chamon,chamon2} by letting $\Gamma \rightarrow \Gamma e^{-i \omega_0 t}$, where $\omega_0=\omega_J=g e V/2 \hbar$.
The tunnelling current operator is $I_t=\frac{1}{i \hbar} [N_R, H]=-\frac{1}{i \hbar} [N_L,H]$ where $N_{R/L}$ are the total charge operators on the $R/L$ edges. Using the commutation relations between the charge and the quasiparticle operators \cite{xgwen}, we find
\be
I_t(t)=i e^* \Gamma \sigma_L^{\dagger}(t,0) \sigma_R(t,0) e^{i \phi(t,0) \sqrt{g}/2} e^{-i \omega_0 t} +h.c.,
\ee
where $e^*= ge/2=e/4$. The fluctuations (shot noise) in the tunnelling current can be written as
\be
S(\omega)=\int_t S(t) e^{i \omega t},
\ee
where
\be
S(t)=\frac{1}{2}\{I_t(t), I_t(0)\}.
\ee
In the next section we will find the expectation value of the tunnelling current operator and of the shot noise up to the first non-zero order in perturbation theory in the tunnelling amplitude $\Gamma$.

\section{Non-equilibrium shot noise and tunnelling current to first non-zero order in perturbation theory}

We note that our problem is non-equilibrium, so that we should use the Keldysh non-equilibrium formalism \cite{Keldysh}.
However, for the zeroth and first order perturbation theory, there is no difference between the equilibrium and non-equilibrium formalisms \cite{chamon}, so in this section we will make use of the equilibrium formalism to study the expectation values of the current and shot noise to first non-zero order in perturbation theory. Thus
\be
\langle I_t\rangle = \frac{1}{\cal Z} \int I_t e^{i S} ,
\ee
where $S$ is the total action $S=S_0+\delta S=\int dt {\cal L}_0+\int dt \delta {\cal L}$, and ${\cal Z}$ is the partition function. We can expand the exponential to first order in the tunnelling to obtain
\be
\langle I_t\rangle =i \int dt' \langle I_t(t) \delta {\cal L}(t')\rangle_0,
\ee
where $\langle \rangle_0$ denotes taking expectation values with respect to the non-perturbed action $S_0$.

Since the expectation value of the current $I_t$ is time independent, we can set $t=0$. Also, one cannot distinguish between $\sigma_{R/L}^{\dagger}$ and $\sigma_{R/L}$, so we will 
write our result in terms of the expectation value of the operator $\sigma=\sigma_R \sigma_L$. Moreover, since $x=0$ for all operators, we will drop the spatial index, and we will only write down the time index. Also, all expectation value symbols on the right hand side refer now to the unperturbed action. The expectation value for the current becomes
\begin{widetext}
\ba
\langle I_t \rangle &=& e^*  |\Gamma|^2 \int dt'   \langle
\sigma(0) \sigma(t')\rangle [e^{i \omega_0 t'} 
\langle e^{i \sqrt{g} \phi(0)/2} e^{-i \sqrt{g}\phi(t')/2}\rangle - e^{-i \omega_0 t'} 
\langle e^{-i \sqrt{g} \phi(0)/2} e^{i \sqrt{g}\phi(t')/2}\rangle ].
\ea
\end{widetext}
We note that $\langle e^{\pm i \sqrt{g} \phi(t)/2} e^{\mp i \sqrt{g}\phi(0)/2} \rangle=e^{g \langle \phi(t) \phi(0)\rangle/4}$ and \cite{chamon} that $\langle \phi(t) \phi(0) \rangle=-2 \ln (\epsilon+i t)$, where $\epsilon$ is a short time/high energy cutoff. Also, we know from CFT \cite{ginsparg}, that $\langle\sigma(z) \sigma(z')\rangle=|z-z'|^{-\delta_\sigma}$, where $\delta_\sigma=1/4$, and $z=\tau+i x$. This yields by analytical continuation $\tau \rightarrow i t+\epsilon$, $\langle\sigma(t) \sigma(0)\rangle=(\epsilon+i t)^{-\delta_\sigma}$.

Putting all the factors together and performing the integrals over time one obtains
\be
\langle I_t \rangle=\frac{2 \pi}{\Gamma(\delta)} e^* |\Gamma|^2 \omega_0^{\delta-1},
\ee
where $\Gamma(x)$ is the Euler Gamma function, $\delta=\delta_{\sigma}+\delta_{\phi}=\delta_{\sigma}+g/2=1/2$, $e^*=g e/2=e/4$, and $\omega_0=g e V/2 \hbar$ is the Josephson frequency corresponding to the applied voltage $V$.

Note that the total current injected in the sample through lead A due to the applied voltage $V$ is $I_0=(g/2) (e^2/h) V$, which is splitting between the leads B and C, $\langle I_t \rangle$ being the tunnelling current picked up in lead B. Obviously, in the absence of tunnelling ($\Gamma=0$) the tunnelling current is zero and the currents in leads A and C are equal, and equal to $I_0$.

Similarly one can calculate the shot noise in the tunnelling current. Up to the first non-zero order in perturbation theory this is:
\ba
\langle S(t) \rangle&&=\frac{1}{2}\langle I(t) I(0)+I(0) I(t)\rangle \nonumber \\
=&&{e^*}^2 |\Gamma|^2  \cos(\omega_0 t) \langle \sigma(t) \sigma(0) \rangle e^{g \langle \phi(t) \phi(0) \rangle/4} +(t \rightarrow -t) \nonumber \\
=&& {e^*}^2 |\Gamma|^2  \cos(\omega_0 t) \big[\frac{1}{(\epsilon+i t)^{\delta}}+\frac{1}{(\epsilon-i t)^{\delta}} \big].
\ea
We note that in the limit of $\epsilon\rightarrow 0$, we can rewrite the expression for $\langle S(t) \rangle$ as follows.
\ba
\langle S(t) \rangle &&={e^*}^2 |\Gamma|^2  \cos(\omega_0 t) \frac{1}{|t|^\delta} [e^{i \pi \delta sgn(t)/2}+e^{-i \pi \delta sgn(t)/2}] \nonumber \\
&&= 2{e^*}^2 |\Gamma|^2 \cos(\omega_0 t) \frac{1}{|t|^\delta} \cos(\pi \delta /2).
\ea
So $\langle S(t) \rangle$, even to first non-zero order in perturbation theory incorporates in it information about both the fractional charge and the fractional statistics of the quasiparticles. For example, $e^*= ge/2$ is a
measure of the fractional charge of the tunnelling quasiparticles. The factor
$\cos(\pi \delta/2)$, where $\delta =g/2+\delta_{\sigma}$ is a measure of the statistical angle, which incorporates combined information about the charge mode through $ g /2$, as well as about the neutral Ising mode through $\delta_{\sigma}$. Also $\langle S(t) \rangle$ decays as a power law in time, with a coefficient $\delta$. Moreover, it oscillates with a periodicity $\omega_0= g e V/2 \hbar$, the Josephson frequency, which  incorporates information  about the fractional charge  $ge/2$. However, no direct  information is retrieved about the non-abelian $\sigma$ Ising mode, since its non-abelian properties only appear for fourth or higher order correlation functions of the $\sigma$ operators\cite{ginsparg}.

Note that for the simple case of tunnelling between the edge states of a Laughlin quantum Hall liquid with filling fraction $\nu$, our results change such that $\delta\rightarrow \delta_l = 2 \nu$, and $e^*\rightarrow e^*_l=\nu e$.

We can also study the frequency dependent noise $\langle S(\omega) \rangle=\int_t e^{i \omega t} \langle S(t) \rangle $. By taking a non-zero cutoff $\epsilon$, and performing the integral over $t$ exactly, one finds that
\ba
 \langle S(\omega) \rangle =&&\frac{\pi}{\Gamma(\delta)}{e^*}^2 |\Gamma|^2 [|\omega-\omega_0|^{\delta-1}+  |\omega+\omega_0|^{\delta-1}] \nonumber \\
=&&\frac{e^{*}}{2} I_t [|1-\omega/\omega_0|^{\delta-1}+  |1+\omega/\omega_0|^{\delta-1}].
\ea
We note that the ratio $\langle S(\omega)\rangle/\langle I_t \rangle$ is a universal function, independent of the strength of the tunnelling $\Gamma$. Moreover, at zero frequency $\langle S(\omega)\rangle /\langle I_t \rangle=e^*$, from where a precise measurements of the fractional charge can be performed, and indeed has been done\cite{shotexp} for the simpler case of a  Laughlin state with $e^*=e/3$. For our situation the fractional charge $e^*=g e/2 =e/4$ of the tunnelling quasiparticles can be extracted. The last feature we note is the presence of the singularity in frequency at the Josephson frequency $\omega=\omega_0$. 

The quantity $\langle S(\omega) \rangle$ seems to yield less information than $\langle S(t) \rangle$, however, it allows for a clean measurement of various quantities such as the fractional charge and the Josephson frequency, independently of the magnitude of the tunnelling coefficient $\Gamma$ or of other experimental factors.
We note that at this order in perturbation theory, it is impossible to
extract direct information about the statistics of the non-abelian Ising mode. Since we are interested in the non-abelian characteristics of the Pfaffian state, we will go to the next order in perturbation theory, from where more information about the non-abelian character of the quasiparticles can be extracted.

We should note that all our calculations till now are performed at zero temperature. The generalization to finite temperature is straightforward, but at this order in perturbation theory the finite temperature will not provide any novel information, so we will not study this situation until later on in the following section.

\section{Higher order perturbative corrections to shot noise}

In the previous section we computed the tunnelling current and the shot noise up to the first non-zero order in perturbation theory. In this section we will study the next order correction. Before starting the calculation, a few comments are in order.  
First, we need to note that to this perturbative order, one also needs to subtract the noncorrelated piece $\langle I_t(t) \rangle \langle I_t(0) \rangle=\langle I_t\rangle ^2$ from the expectation value of  the shot noise $\langle S(t) \rangle$. Subtracting this contribution has no effect to the order $|\Gamma|^2$, as $\langle I_t\rangle^2$ is of order $ |\Gamma|^4$, but needs to be considered when we compute the shot noise to the order  $|\Gamma|^4$.
We expect that $\langle S(t) \rangle -\langle I_t \rangle^2$ goes to zero when $t$ is large, as the current should be uncorrelated with itself for large time separations. 
Moreover, the tunnelling current $\langle I_t \rangle$ is independent of time, so the correction to $\langle S(t) \rangle$ will be a mere constant.
In principle this correction can be determined easily from our calculations, but since it may slightly depend on various cutoffs, etc., we will account for it by automatically renormalizing $\langle S(t)\rangle$ such that it goes to zero for large $t$.
 
The second thing we need to note is that we are usually interested not in the actual $\langle S(\omega)\rangle $, but in the ratio $\langle S(\omega)\rangle /\langle I_t \rangle$. If we compute $\langle S(\omega)\rangle$ up to $|\Gamma|^4$, in order to compute the ratio accurately to order $|\Gamma|^4$, then we should also carefully compute $\langle I_t \rangle$ to order $|\Gamma|^4$. However, since $\langle I_t \rangle$ is a constant with respect to the frequency $\omega$, this will only renormalize the ratio between the first and the second order corrections in $\langle S(\omega)\rangle /\langle I_t\rangle$, and thus could be easily accounted by a slight renormalization in the tunnelling coefficient $\Gamma$.  For these reasons, in this paper we will neglect these higher order corrections to $\langle I_t \rangle$.                                   

However, we should note that $\langle I_t \rangle $ depends on the applied voltage, so that if we want to consider the dependence of $\langle S(\omega)\rangle /\langle I_t\rangle$ at a fixed frequency $\omega$ on the applied voltage, one needs to take also into account the corrections of order $|\Gamma|^4$ to $\langle I_t \rangle$. It actually may be a good exercise with interesting consequences on its own, to compute these higher order corrections in $\langle I_t \rangle $ and see what, if any, are the consequences of the non-abelian and fractional statistics for the tunnelling current.

\begin{figure}[h]
\begin{center}
\includegraphics[width=2in]{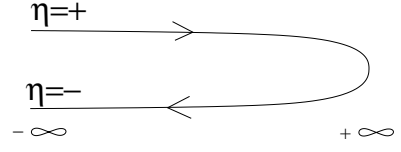}
\end{center}
\vspace{-0.15in} \caption{The two branches $\eta=\pm$ of the Keldysh contour} \label{fig01}
\end{figure}

In order to compute the next order in perturbation theory for $\langle S(t) \rangle$, it does no longer suffice to resort to an equilibrium approach, we need to use a Schwinger- Keldysh non-equilibrium approach\cite{Keldysh}. We thus double the integration contour such that our time integrals extend from $-\infty$ to $\infty$ (the $\eta=+$ part of the contour), and back from $\infty$ to $-\infty$ (the $\eta=+$ part of the contour) (see Fig.\ref{fig01}). We thus introduce a new index $\eta$ to our operators which describes the branch of the contour on which a specific operator sits.
\vspace{.05in}
In terms of the new operators, we can write
\be
S(t)=\frac{1}{2} [I^+(t) I^-(0)+I^+(0) I^-(t) ]
\ee
and
\be \langle S(t) \rangle=\frac{1}{2}\sum_{\eta=\pm} \langle T_K I^{\eta}(t) I^{-\eta}(0) e^{i \int_K dt \delta {\cal  L}(t)}\rangle_0,
\label{k1}
\ee
where again we suppressed the spatial indices and the $\langle \rangle_0$ denotes expectation value with respect to the free action $\int_K dt \int d^3 x {\cal L}_0$. The symbol $T_K$ denotes time ordering along the Keldysh contour described above and depicted in Fig. \ref{fig01}, and $\int_K$ denotes a time integral along the Keldysh contour, $\int_K dt=\sum_{\eta=\pm} \eta \int_{-\infty}^{\infty} dt \equiv \sum_{\eta=\pm} \eta \int  dt$.
 Expanding the exponentials in Eq.(\ref{k1}) we can write the second non-zero correction to $S(t)$ as
\ba 
\langle S(t) \rangle_1=&&-\frac{1}{2}\sum_{\eta=\pm} \langle T_K \int_K dt_1 \int_K dt_2  \times \nonumber \\ \times && I^{\eta}(t) I^{-\eta}(0) \delta {\cal  L}(t_1) \delta {\cal  L}(t_2) \rangle_0.
\ea

The details of computing $\langle S(t) \rangle_1$ are given in Appendix A.  
We obtain
\begin{widetext}
\begin{eqnarray}
\langle S(t) \rangle_1=-2 {e^*}^2 |\Gamma|^4\Big( &\int_{t_1,t_2}^{t_1<0<t_2<t} &
(f_{tt_1}^{\delta_\phi}-f_{t0}^{\delta_\phi})\{\tilde{f}_{tt_2}^{\delta_\sigma}\cos \pi 
(\delta_\phi-\delta_\sigma)
+\tilde{f}_{t0}^{\delta_\sigma} [\cos \pi   
(\delta_\phi+\delta_\sigma)-2]\}
\nonumber
\\&&
+f_{tt_2}^{\delta_\phi}\{\tilde{f}_{tt_2}^{\delta_\sigma}\cos \pi 
(\delta_\phi+\delta_\sigma)
+\tilde{f}_{t0}^{\delta_\sigma} [\cos \pi   
(\delta_\phi+\delta_\sigma)-1-\cos 2 \pi \delta_\phi]\}
\nonumber \\
+&\int_{t_1,t_2}^{t_1<t_2<0<t}&
f_{tt_1}^{\delta_\phi} [\tilde{f}_{tt_2}^{\delta_\sigma}\cos \pi 
(\delta_\phi+\delta_\sigma)
-\tilde{f}_{t0}^{\delta_\sigma} \cos \pi   
(\delta_\phi-\delta_\sigma)]
\nonumber
\\&&
+f_{tt_2}^{\delta_\phi}[\tilde{f}_{tt_2}^{\delta_\sigma}\cos \pi 
(\delta_\phi+\delta_\sigma)
-\tilde{f}_{t0}^{\delta_\sigma} \cos \pi   
(\delta_\phi+\delta_\sigma)]
\nonumber
\\&&
+f_{t0}^{\delta_\phi}\{\tilde{f}_{t0}^{\delta_\sigma}\cos \pi 
(\delta_\phi+\delta_\sigma)
+\tilde{f}_{tt_2}^{\delta_\sigma} [1-\cos \pi   
(\delta_\phi+\delta_\sigma)-\cos 2 \pi \delta_\phi]\}
\nonumber \\
+&\int_{t_1,t_2}^{0<t_1<t_2<t}&
(f_{tt_1}^{\delta_\phi}-f_{t0}^{\delta_\phi})[\tilde{f}_{tt_1}^{\delta_\sigma}\cos \pi 
(\delta_\phi+\delta_\sigma)
-\tilde{f}_{tt_2}^{\delta_\sigma} \cos \pi   
(\delta_\phi-\delta_\sigma)]
\nonumber
\\&&
-f_{tt_2}^{\delta_\phi}\{\tilde{f}_{tt_2}^{\delta_\sigma}\cos \pi 
(\delta_\phi+\delta_\sigma)
+\tilde{f}_{tt_1}^{\delta_\sigma} [\cos 2 \pi \delta_\phi-1-\cos \pi   
(\delta_\phi+\delta_\sigma)]\}\Big)
\label{res0}
\end{eqnarray}
where
\ba
 f_{t 0}^{\mu}&=&|t|^\mu |t_1-t_2|^{\mu} |t-t_1|^{-\mu} |t-t_2|^{-\mu}
 |t_1|^{-\mu} |t_2|^{-\mu} \cos[\omega_0(t-t_1-t_2)]
\nonumber \\
f_{t t_1}^{\mu}&=&|t-t_1|^\mu|t_2|^{\mu} |t|^{-\mu} |t-t_2|^{-\mu}
|t_1-t_2|^{-\mu} |t_1|^{-\mu}   \cos[\omega_0(t+t_1-t_2)]
\nonumber \\
f_{t t_2}^{\mu}&=&|t-t_2|^\mu |t_1|^{\mu} |t-t_1|^{-\mu} |t|^{-\mu}
|t_1-t_2|^{-\mu} |t_2|^{-\mu}   \cos [\omega_0(t+t_2-t_1)],
\ea    
$\tilde{f}^{\mu}_{t 0, t_1,t_2}=f^{\mu}_{t 0 t_1 t_2}(\omega_0\rightarrow 0)$, and for the Pfaffian state $\delta_{\sigma}=\delta_{\phi}=1/4$.
This reduces to:
\begin{eqnarray}
\langle S(t) \rangle_1=&&-2 {e^*}^2 |\Gamma|^4
\Big\{ \int_{t_1,t_2}^{t_1<0<t_2<t} 
[(f_{tt_1}^{\delta_\phi}-f_{t0}^{\delta_\phi})
(\tilde{f}_{tt_2}^{\delta_\sigma}
-2 \tilde{f}_{t0}^{\delta_\sigma})
-f_{tt_2}^{\delta_\phi}\tilde{f}_{t0}^{\delta_\sigma}]
\nonumber \\&&
+\int_{t_1,t_2}^{t_1<t_2<0<t}(
f_{t0}^{\delta_\phi}\tilde{f}_{tt_2}^{\delta_\sigma}-
f_{tt_1}^{\delta_\phi}\tilde{f}_{t0}^{\delta_\sigma})
+\int_{t_1,t_2}^{0<t_1<t_2<t}(
f_{t0}^{\delta_\phi}\tilde{f}_{tt_2}^{\delta_\sigma}-
f_{tt_1}^{\delta_\phi}\tilde{f}_{tt_2}^{\delta_\sigma}
+f_{tt_2}^{\delta_\phi}\tilde{f}_{tt_1}^{\delta_\sigma})\Big\}
\end{eqnarray}

For comparison, we also provide the similar result for the case of a Laughlin state with filling factor $\nu$ obtained from the more 
general Pfaffian result by setting $\delta_\phi\rightarrow 2 \nu$, and    
$\delta_\sigma \rightarrow 0$.
\begin{eqnarray}
\langle S(t) \rangle_1= &&-2 {e^*}^2 |\Gamma|^4 \big\{ \int_{t_1,t_2}^{t1_<0<t_2<t} [2 f_{t 0}^{\delta_l}(1-\cos \pi \delta_l)+2 f_{t t_1}^{\delta_l}(\cos \pi \delta_l-1)+f_{t t_2}^{\delta_l}(2 \cos \pi \delta_l-1-\cos 2 \pi \delta_l)] \nonumber \\&& +\int_{t_1,t_2}^{t1_<t_2<0<t} f_{t0}^{\delta_l}(1-\cos 2 \pi \delta_l) 
+\int_{t_1,t_2}^{0<t1_<t_2<t} f_{t t_2}^{\delta_l}(1-\cos 2 \pi \delta_l) \big\},
\end{eqnarray}
\end{widetext}
and $\delta_l=2 \nu$.   

A sketch of the integration domains is presented below:
\vspace{.2in}
\begin{figure}[h]
\begin{center}
\includegraphics[width=2.5in]{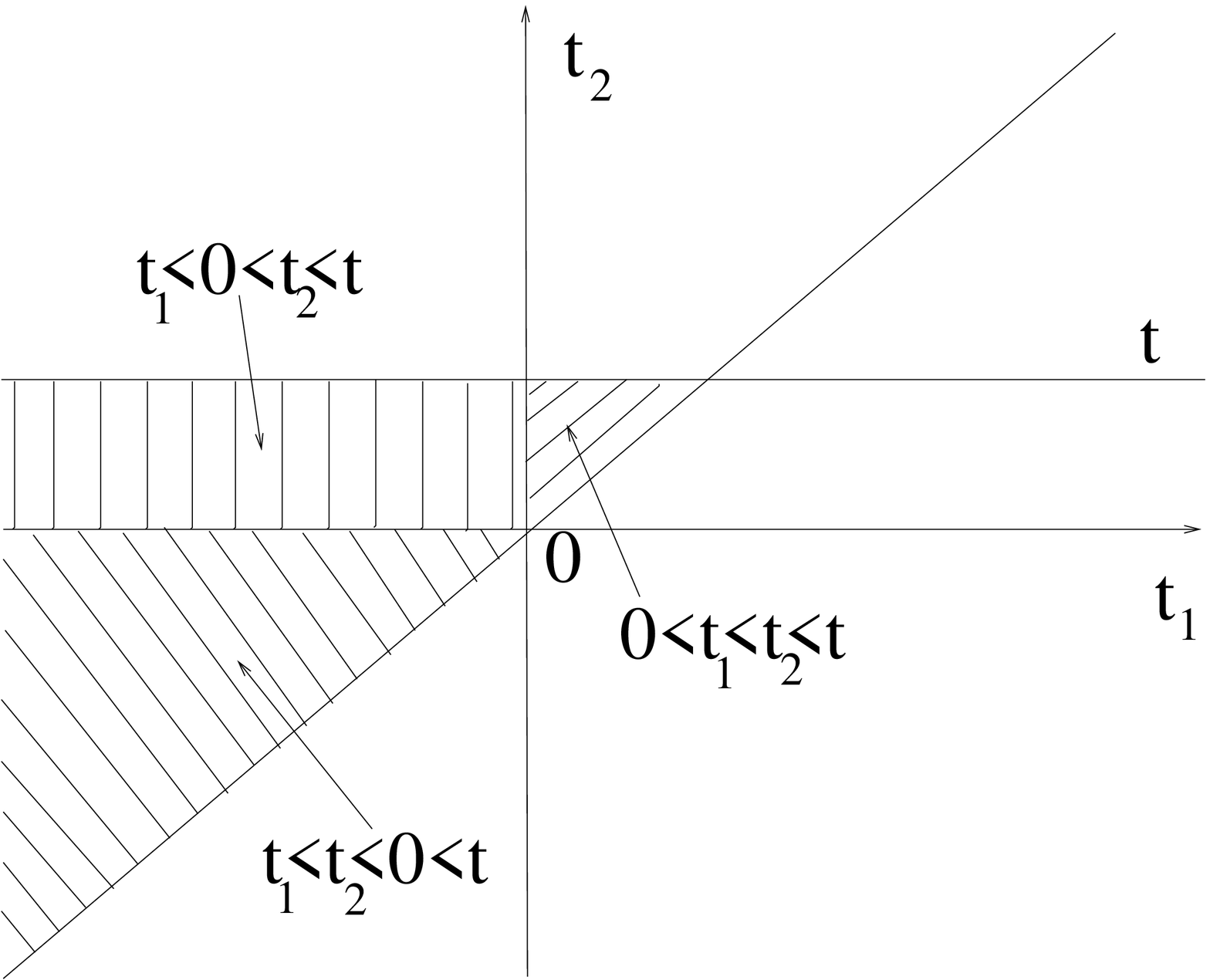}
\end{center}
\vspace{-0.15in} \caption{Intervals of integration} \label{fig03}
\end{figure}      

\section{Discussion and Results}

A few comments about $\langle S(t)\rangle_1$ are in order. Note that we focus only on $t>0$, as $\langle S(t) \rangle$ is symmetric under $t\rightarrow -t$. Also, when we study it numerically we need to subtract the corresponding constant  $\langle I_t \rangle^2$, such that  $\langle S(t)\rangle_1$ goes to zero for large time separations. The result described above is for zero temperature, and can be translated to finite temperatures by a conformal transformation under which $|t|$ becomes $\sinh (|t| \pi T)/\pi T$.

We also note that we can expand $\cos[\omega_0(t+\epsilon_1 t_1 +\epsilon_2 t_2)]$ as $\cos(\omega_0 t)\cos[\omega_0(\epsilon_1 t_1 +\epsilon_2 t_2)]-\sin(\omega_0 t)\sin[\omega_0(\epsilon_1 t_1 +\epsilon_2 t_2)]$, where $\epsilon_{1/2}=\pm 1$. Thus, after performing the integrals over $t_1$ and $t_2$, $\langle S(t)\rangle_1$  can be written as $\cos (\omega_0 t) f_1(t) +\sin (\omega_0 t) f_2(t)$, where $f_1$ and $f_2$ will be oscillatory decaying functions. The nice thing about this way of writing $\langle S(t)\rangle_1$ is that the contributions form the terms proportional to $\cos(\omega_0 t)$, and $\sin(\omega_0)t$ in  $\langle S(t)\rangle_1$  can be separated. In general the first type of terms will yield a Lorentzian peaked about $\omega_0$ when Fourier transformed, while the second type of terms will yield a derivative of a Lorentzian type curve, also centered about the Josephson frequency $\omega_0$. 

Thus, we expect that, when Fourier transformed, each term in equation (\ref{res0}) yields a superposition of Lorentzians and Lorentzian derivatives centered about the Josephson frequency $\omega_0$. Of course, the magnitude and the sign of each term depend on the interval of integration, as well as on the corresponding statistical factors. Also, the sharpness of each resulting Lorentzian peak/peak derivative will be determined by the temperature, as well as by the exponent of the corresponding power law. The exact shape of $\langle S(\omega)\rangle_1$ will be thus determined by the interplay of many factors, and in general cannot be guessed without a detailed numerical evaluation of each term.

Nevertheless, there a few observations one can make. 
The first is that in the case of a Laughlin state all the power laws involved in $\langle S(t)\rangle_1$ come as $f_{t0,t t_1,t t_2}^{\delta_l}$, while in the Pfaffian state some terms come as $f_{t0,tt_1,tt_2}^{\delta_\sigma+\delta_\phi}$, while others come as  combinations of $f^{\delta_{\phi}}_{t 0,t_1,t_2}$ and
$\tilde{f}^{\delta_{\sigma}}_{t 0,t_1,t_2}$. A similar observation can also me made about the statistical factors, in the Laughlin state all the statistical factors involve $\delta_l$, while in  the Pfaffian state they will involve $\delta_{\phi}+\delta_{\sigma}$ and $\delta_{\phi}-\delta_{\sigma}$ .
The sources of these difference are the non-abelian
characteristics of the four point correlation function of the $\sigma$ operators. Specifically, as described in detail in Appendices A and B, the four point correlation function for the $\sigma$ operators has different forms, depending on the order of the four times at which the correlation function is evaluated. Thus for each time ordering of $t_1, t_2, 0$ and $t$, and thus for each domain of integration, each of the term to be integrated will have a different form. 

This will have a double effect, one on the relative coefficients, and one on the shape of each term. We can compare the relative magnitudes of each term. 
We note that for the Pfaffian case the integral is dominated by the
terms involving $f_{t t_2}^{\delta_\phi}$ and $f_{t t_1}^{\delta_\phi}$
on the interval $0<t_1<t_2<t$. In particular, the term $f_{t
t_1}^{\delta_\phi}$ is responsible for the differences between  the   
Laughlin and the Pfaffian cases and it vanishes in the Laughlin state,
where the result is dominated by the term $f_{t t_2}^{\delta_\phi}$.

The value of $\langle S(\omega) \rangle_1$, for a Pfaffian and Laughlin state ($\nu=1/5$) are plotted below. We chose the values of the applied potential such that $\omega_0=1$. Also we chose the temperature to be $T=0.1$, and tunnelling strength $|\Gamma|^2=0.5\%$ (Pfaffian), and $|\Gamma|^2=0.8 \%$ (Laughlin). The numerical integrals over $t_1$ and $t_2$ are done using Mathematica. The singularities at $t=t_{1/2},0$, and $t_1=t_2$ are avoided by stopping the integral at some $\delta_t=0.001$ away from the singularities (this is equivalent to imposing a high energy hard cutoff). We could have imposed a soft cutoff as in the previous section by substituting $ i t \rightarrow \epsilon + i t$ but the hard cutoff procedure is more transparent for the observation of statistics, and it is also easier to implement numerically. Also, note the rounding of the Josephson singularities due to finite temperature effects.
\begin{figure}[h]
\begin{center}
\includegraphics[width=2.5in]{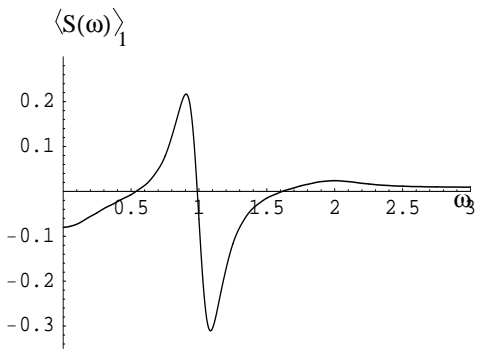}
\end{center}
\vspace{0.15in} \caption{The second order correction to the frequency dependent shot noise $\langle S(\omega) \rangle_1$ (in units of $e^* \langle I_t \rangle$), as a function of frequency (in units of $\omega_0$) in the Pfaffian state. We set $T=0.1$, and $|\Gamma|^2=0.5\%$} \label{fig2}
\end{figure}  
\begin{figure}[h]
\begin{center}
\includegraphics[width=2.5in]{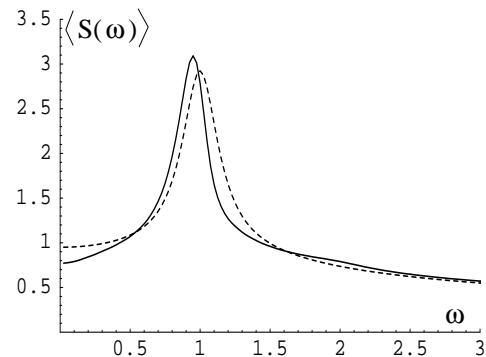}
\end{center}
\vspace{0.15in} \caption{The total shot noise $\langle S(\omega) \rangle$ (in units of $e^* \langle I_t \rangle$) as a function of frequency (in units of $\omega_0$) in the Pfaffian state. The dotted line is the first order contribution to shot noise. The solid line is the shot noise coming from both the first and the second order terms. 
Note the shifting and the
enhancing of the peak at the Josephson frequency as a result of including the higher order term, and the low frequency reduction in the shot noise. We set $T=0.1$, and $|\Gamma|^2=0.5\%$} \label{fig3}
\end{figure}    
 \begin{figure}[h]
\begin{center}
\includegraphics[width=2.5in]{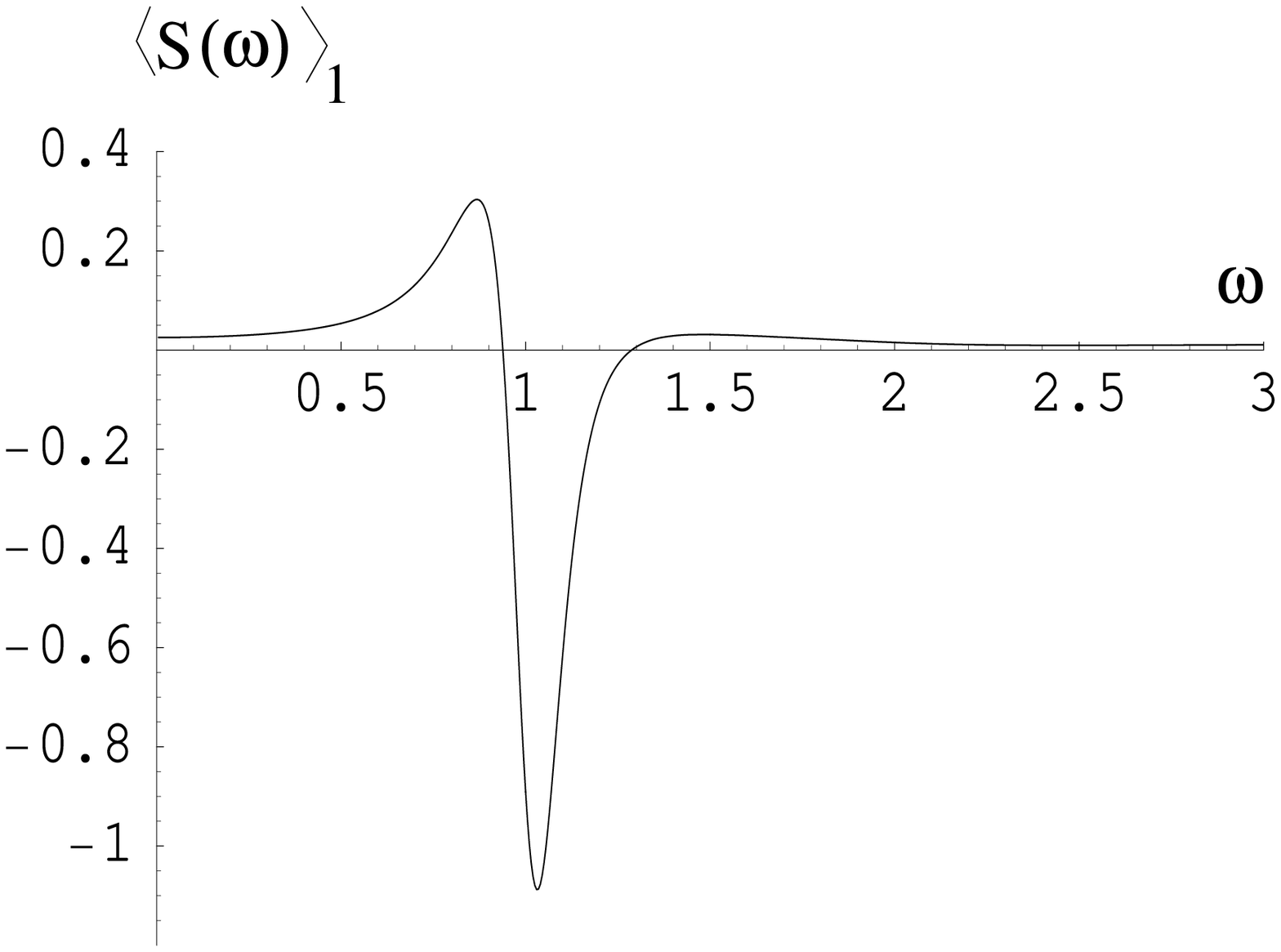}
\end{center}
\vspace{0.15in} \caption{The second order correction to the frequency dependent shot noise $\langle S(\omega) \rangle_1$ (in units of $e^*_l \langle I_t \rangle$), as a function of frequency (in units of $\omega_0$) in the Laughlin state. We set $T=0.1$, and $|\Gamma|^2=0.8\%$} \label{fig4}
\end{figure}  
\begin{figure} [h]
\begin{center}
\includegraphics[width=2.5in]{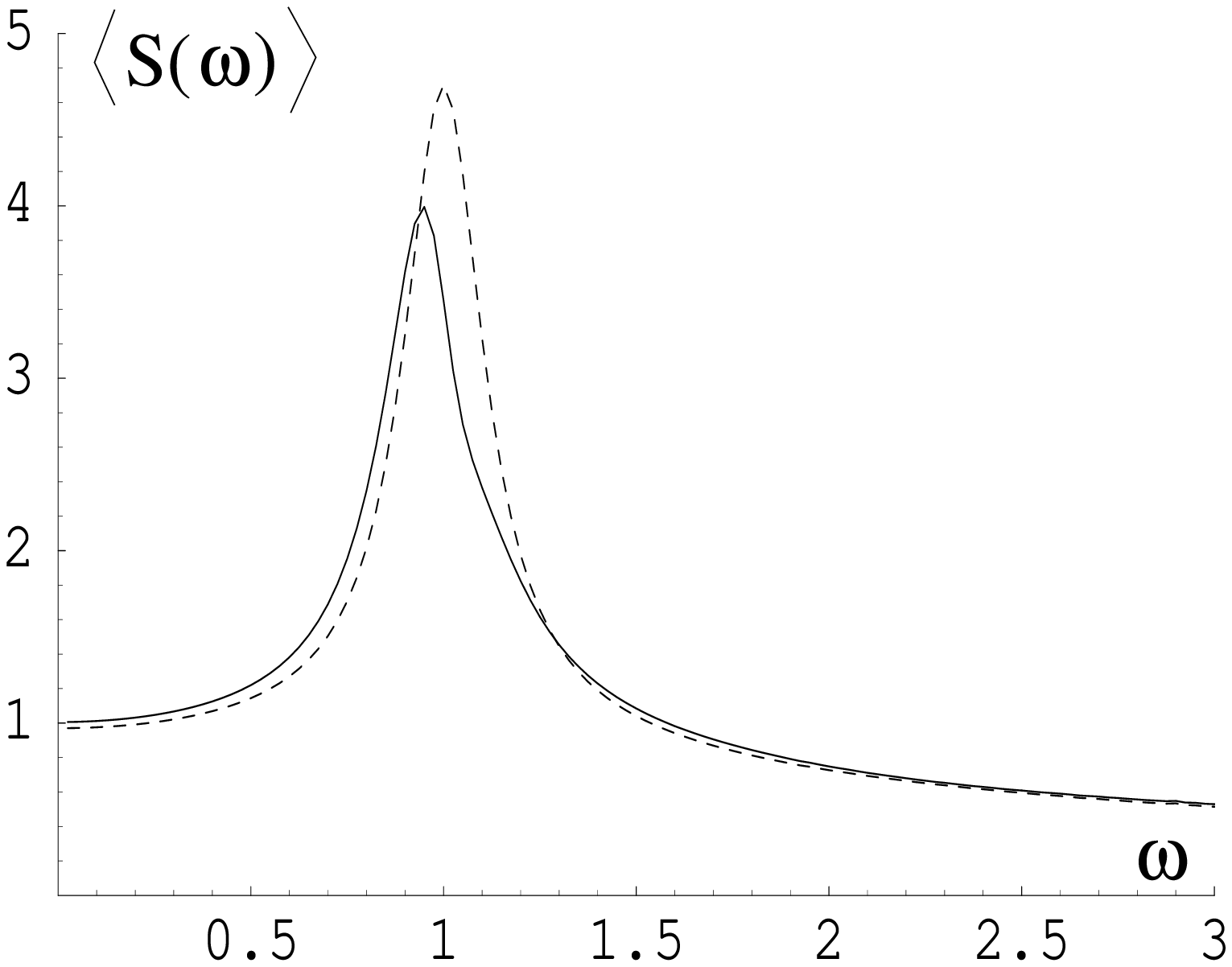}
\end{center}
\vspace{0.15in} \caption{The total shot noise $\langle S(\omega) \rangle$ (in units of $e^*_l \langle I_t \rangle$) as a function of frequency (in units of $\omega_0$) in the Laughlin state. The dotted line is the first order contribution to shot noise. The solid line is the shot noise coming from both the first and the second order terms. Note the shifting and   
the flattening of the peak at the Josephson frequency as a result of including the higher order term. We set $T=0.1$, and $|\Gamma|^2=0.8\%$} 
\label{fig5}
\end{figure}    

We note that the effect of the higher order correction in the noise is an enhancing and a shifting of the singularity towards lower
frequncies in the Pfaffian state and a flattening of the singularity in
the Laughlin state. Also we note a decrease in the shot noise at zero and small frequencies, such that the ratio $\langle S(\omega) \rangle/\langle I_t \rangle$ decreases slightly in the Pfaffian state and in the Laughlin state with
$\nu=1/3$ and increases slightly for the Laughlin state with $\nu=1/5$. This effect appears to be more significant in the Pfaffian state in the sense that
a tunneling strength $\Gamma$ which produces comparable shifts of the Josephson frequency leads to a much larger suppression of the zero-frequency
shot noise in the case of the Pfaffian.

This effect is not as dramatic as the underlying physics; once it is Fourier transformed and
inserted into the expression for the noise, the structure in the quasiparticle four-point function is masked. Hence, it might also be worth analyzing other setups, see for example the setups in Ref.~\onlinecite{eunah,shotproposals}. A very similar calculation for abelian  Laughlin and Jain states has been performed in a three terminal geometry\cite{eunah}. The authors conclude that the dependence of shot noise in their setup also included Lorentzian peaks and peak derivatives, yielding contributions similar to our findings, and which could distinguish between the Laughlin and Jain states. This setup is harder to achieve experimentally, but it has the advantage that, while in our setup the relevant non-abelian physics comes in only as a second order correction, in a three terminal setup, it may be evident in the first non-zero order, which would make it easier to observe experimentally.
In the next section, we make a few observations about the formalism needed for such a calculation.

To sharpen the distinction between the Pfaffian state and other quantum Hall states (especially
Abelian ones), it would be useful to repeat our calculation for the Jain sequence, the (3,3,1)
state, etc., and compare these results with those given above.

\section{Multi-Terminal Setup}

\begin{figure}[t]
\begin{center}
\includegraphics[width=2.75in]{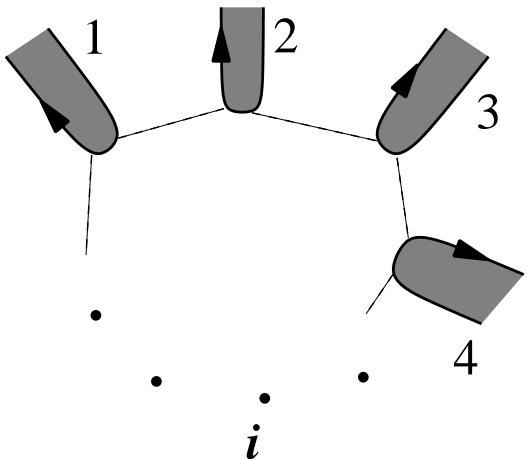}
\end{center}
\caption{A multi-lead setup. We treat the edge theory operators at the different
leads as independent, but must introduce Klein factors into our formalism in order
to correctly account for quasiparticle statistics.}
\label{fig:multi-lead}
\end{figure}

Since the effects of non-Abelian statistics first appear
in fourth-order correlation functions, it would be advantageous to look
at experimental setups in which the leading processes occur at this order.
Consider the setup depicted in figure \ref{fig:multi-lead}.
In order to describe tunneling between the different leads, we
introduce multiple chiral Pfaffian edges $\sigma_i$, $\psi_i$, $\phi_i$,
with $i=1,2,\ldots,n$. (These fields are all chiral, so we do not bother adding
a subscript $R$.)
Naively, the tunneling operator
between edges $i$ and $j$ takes the form:
\be
H_{\rm tun}^{i,j}=-\Gamma \sigma_i
\sigma_j e^{i\left({\phi_i}-{\phi_j}\right) \sqrt{g}/2}+h.c.
\label{eqn:multi-lead-naive}
\ee
However, this is not quite right because it treats the different edges as
completely independent. In reality, they are not quite independent
because quasiparticles at the different edges must still have
(non-Abelian!) braiding statistics. This can be accommodated by
modifying (\ref{eqn:multi-lead-naive}) to
\be
H_{\rm tun}^{i,j}=-\Gamma {F_i^\dagger}{F_j}\:{\tau_i}{\tau_j}\: \sigma_i
\sigma_j e^{i\left({\phi_i}-{\phi_j}\right) \sqrt{g}/2}+h.c.
\label{eqn:multi-lead-Klein}
\ee
by introducing `Klein factors' $F_i$ and $\tau_i$, $i=1,2,\ldots,n$.
These operators satisfy commutation relations which
are `reverse engineered' to give the correct quasiparticle
statistics. The $F_i$s account for the Abelian part
and satisfy the relations \cite{Guyon02}:
\begin{eqnarray}
{F_i^\dagger}\,{F_i} &=& 1\cr
{F_i}\,{F_j} &=& e^{\frac{i \pi g}{4}\,p_{ij}}\,{F_j}\,{F_i}
\end{eqnarray}
where $p_{ij}=0,\pm 1$ are chosen to ensure that tunneling
operators commute if their tunneling paths do not intersect
but acquire a phase $\pm i \pi g/4$ upon commutation
if the tunneling paths have intersection number $\pm 1$.
(Remember that $g=1/2$ for a Pfaffian state at
$\nu=5/2$.)

These relations were obtained by requiring that $H_{\rm tun}^{i,j}$
and $H_{\rm tun}^{k,l}$ have the same properties as anyonic
quasiparticle trajectories (or Wilson lines). In the non-Abelian
case, this generalizes to a skein relation. This skein relation
can only be satisfied with matrices, from which the non-Abelian
nature of the statistics follows. Following the logic used in the ABelian
case, but now generalized to this more complicated situation,
we suggest that the correct statistics can be implemented
by making the $\tau_i$s satisfy the following relations.
\begin{equation}
{\tau_i^2} = 1
\end{equation}
To write down the commutation relations between different
$\tau_i$s, we group the leads into pairs.
This can be done arbitrarily but, for
convenience, we group them as $(1,2)$, $(3,4)$, $\ldots$, $(2j-1,2j)$, $\ldots$. Each pair of leads has a gapless fermionic level
associated with it. The commutation relation between
two $\tau_i$s from the same pair is:
\begin{eqnarray}
\label{eqn:non-Ab-Klein}
\tau_{2j-1} \tau_{2j} =
e^{\frac{i\pi}{4} \sigma_{j}^z}\:\tau_{2j} \tau_{2j-1}
\end{eqnarray}
When a quasiparticle from one pair of leads hops onto a lead from
a different pair, the two two-level systems are coupled. The
commutation relation between two $\tau_i$s from different
pairs features a $4\times 4$ matrix which acts on these
two two-level systems
\begin{eqnarray}
\tau_{k} \tau_{l} = \frac{1}{\sqrt{2}}
\left(
 \begin{smallmatrix}
  1 & -i{\sigma_x}  \\
  -i{\sigma_x} & 1 \\
\end{smallmatrix}
 \right)^{p_{kl}}
\:\tau_{l} \tau_{k}
\end{eqnarray}
The numbers $p_{kl}$ are the same as for the
Abelian Klein factors $F_i$.

In the two-terminal case, which was our primary interest in this paper,
these Klein factors should, strictly speaking be present, but they have no effect since the factors associated with hopping from lead $1$ to lead $2$ and its conjugate commute, e.g. $({\tau_1} {\tau_2})({\tau_2}{\tau_1})=({\tau_2}{\tau_1})({\tau_1} {\tau_2})=1$.
In the multi-lead case, Klein factors can have a dramatic effect on correlation functions
even in the Abelian case, where their commutation relations take a simpler form
\cite{Guyon02,eunah}. The same is true here.
In particular, we expect that some of the effects which we have found at fourth order in
the two-terminal case will now appear at lowest order for current cross-correlations
between different leads. This will be further investigated elsewhere \cite{next-paper}.

\section{Conclusion}

We computed the shot noise in the tunnelling current between two edges of a quantum Hall fluid in a Pfaffian state. Specifically we analyzed the frequency dependence of the shot noise perturbatively in the tunnelling strength. We focused on the first two non-zero orders in perturbation theory. We found that, while the first non-zero order reveals some interesting physics, it does not encode any information about  the non-abelian nature of the Pfaffian state. We also found that the second order in perturbation theory indeed contains information about the non-abelian physics. The effects of the non-abelian physics are manifest in the shot noise, in that the higher order correction has different features than, for example, a regular Laughlin state.  Specifically, we found that the singularity in the shot noise at the Josephson frequency is shifted towards lower frequency and enhanced in the
Pfaffian state and flattened in the Laughlin state. Also, the ratio of $\langle S(t)\rangle /\langle I_t \rangle$ at zero and small frequencies is decreased in the Pfaffian 
state  and in the Laughlin state with $\nu=1/3$,and increased in the
Laughlin state with $\nu=1/5$. Unfortunately, we cannot say if these features are entirely distinctive for a Pfaffian state and non-abelian statistics, and may appear also in other types of states, e.g. Jain, Halperin, 331.
We hope however that our analysis and results will be a first step towards a thorough analysis of the effects of the non-abelian statistics in various other multi-terminal shot noise experiments, where more definite effects might be easier to isolate, and also will increase the interest to perform the shot noise experiments that would look for these sort of features. We must note that much of the relevant physics should come in at the Josephson frequency $ g e V/2 \hbar$ in the range of $GHz$ to $THz$, for applied voltages of order of microvolts to millivolts, which should be observable experimentally.

\acknowledgments
We would like to thank Eun-Ah Kim, Michael Lawler, and Smitha Vishveshwara for helpful discussions and comments. C.~B.\ and C.~N.\ have been supported by
the ARO under Grant No.~W911NF-04-1-0236.
C.~N.\ has also been supported by the NSF under
Grant No.~DMR-0411800.

\appendix
\begin{widetext}

\section{Perturbative expansion for the current noise}

We will evaluate the second order correction to $S(t)$:
\begin{eqnarray} 
\langle S(t) \rangle_1=-\frac{1}{2}\sum_{\eta=\pm} \langle T_K \int_K dt_1 \int_K dt_2  
I^{\eta}(t) I^{-\eta}(0) \delta {\cal  L}(t_1) \delta {\cal  L}(t_2) \rangle_0.
\end{eqnarray}
Here $\int_K dt=\sum_{\eta=\pm} \eta \int_{-\infty}^{\infty} dt \equiv \sum_{\eta=\pm} \eta \int  dt$. Also we have
\begin{equation}
\delta {\cal L}(t)=-\Gamma \hat{T}(t) e^{-i \omega_0 t}-\Gamma^* \hat{T}^{\dagger} e^{i \omega_0 t}.
\end{equation}
and
\be
I_t(t)=i e^* \Gamma \hat{T}(t) e^{-i \omega_0 t}-i e^* \Gamma \hat{T}^{\dagger}(t) e^{i \omega_0 t},
\ee
where $\hat{T}(t)=\sigma(t) e^{i \phi(t)}$, and $e^*=g e/2$.
Thus we obtain for $S(t)$:
\begin{eqnarray}
\langle S(t) \rangle&=&-\frac{1}{2} (i e^*)^2 |\Gamma|^4 \sum_{\eta, \eta_1, \eta_2} \eta_1 \eta_2  \sum_{\epsilon=\pm 1} \sum_{\epsilon_1=\pm 1} \sum_{\epsilon_2=-1}^{-\epsilon_1} \epsilon_1
\int dt_1 \int dt_2 e^{i \epsilon \omega_0 [t+\epsilon_2 t_1-(1+\epsilon_1+\epsilon_2) t_2]}\times \nonumber \\&& \times \langle T_K \hat{T}^\epsilon(t^\eta)\hat{T}^{\epsilon \epsilon_1}(0^{-\eta}) \hat{T}^{\epsilon \epsilon_2}(t_1^{\eta_1})\hat{T}^{-\epsilon(1+\epsilon_1+\epsilon_2)}(t_2^{\eta_2})\rangle_0 .
\end{eqnarray}

We note that $\langle S(t) \rangle_1$ must be symmetric for $t \rightarrow -t$, thus we are only going to focus on $t>0$. We also note that in this situation, various terms from the above expansion cancels for various ordering of the times $t_1,t_2,0$ and $t$. This is expected, due to causality, i.e., processes in which either one or both of two virtual quasiparticle tunnelling between edges happen after the measurement of $\langle I_t \rangle$ cannot contribute to fluctuations. In other words, only the situations for which at least one of the conditions $t_1<0<t$ and $t_2<0<t$ holds will contribute. Noting the symmetry in exchanging $t_1$ and $t_2$, we conclude that we can write $\langle S(t)\rangle_1$ as a sum of three integrals over three domains of integration in the $(t_1,t_2)$ plane. The three domains of integration are depicted in Fig. \ref{fig04}. The intervals symmetric about the $t_1=t_2$ axis will provide an equal contribution, which is taken into account by an extra factor of $2$.
\vspace{.2in}
\begin{figure}[h]
\begin{center}
\includegraphics[width=2.5in]{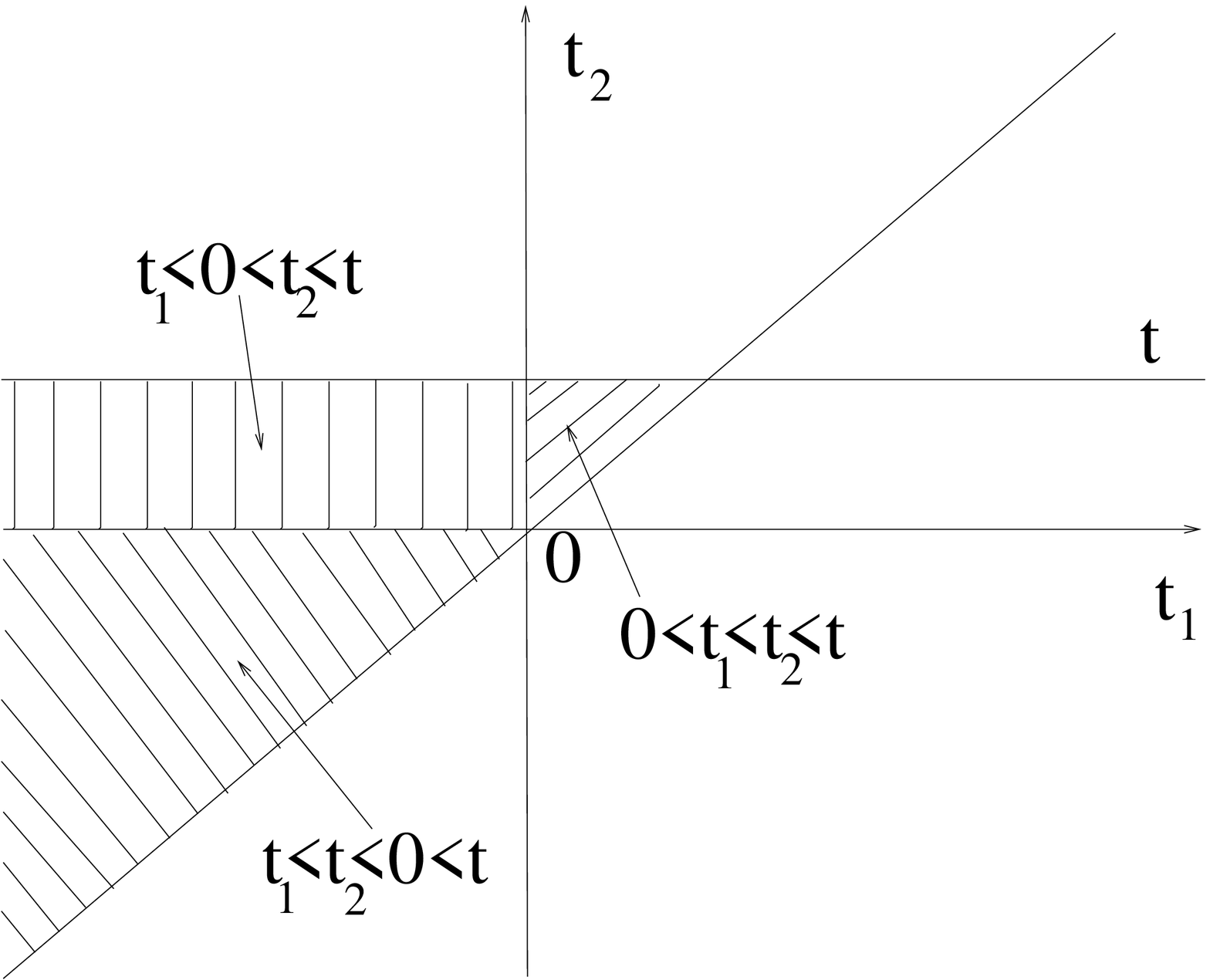}
\end{center}
\vspace{-0.15in} \caption{Intervals of integration} \label{fig04}
\end{figure}
Given the fact that in each individual integration domain the ordering of the four times is well defined we can explicitly evaluate the Keldysh time-ordering operator. We thus obtain
\begin{eqnarray}
\langle S(t) \rangle= &&-{e^*}^2 |\Gamma|^4 \{ \int_{t_1,t_2}^{t1_<0<t_2<t} [F(0,t,t_2,t_1)+F(t,t_2,0,t_1)-F(0,t_2,t,t_1)-F(t_2,t,0,t_1)] \nonumber \\&& +\int_{t_1,t_2}^{t1_<t_2<0<t} [F(0,t,t_2,t_1)+F(t,0,t_2,t_1)-F(t_2,0,t,t_1)-F(t_2,t,0,t_1)] \nonumber \\&& +\int_{t_1,t_2}^{0<t1_<t_2<t} [F(0,t,t_2,t_1)+F(t,t_2,t_1,0)-F(0,t_2,t,t_1)-F(t_2,t,t_1,0)] \}+h.c.
\end{eqnarray}
where 
\begin{eqnarray}
F(\alpha_1,\alpha_2,\alpha_3,\alpha_4)=&&\langle \sigma(\alpha_1) \sigma(\alpha_2) \sigma(\alpha_3) \sigma(\alpha_4) \rangle \{F_{t t_1}(\alpha_1,\alpha_2,\alpha_3,\alpha_4) \cos[\omega_0(t+t_1-t_2)]
\nonumber \\&&+F_{t t_2}(\alpha_1,\alpha_2,\alpha_3,\alpha_4) \cos[\omega_0(t+t_2-t_1)]
\nonumber \\&&
-F_{t 0}(\alpha_1,\alpha_2,\alpha_3,\alpha_4) \cos[\omega_0(t-t_2-t_1)]\}.
\end{eqnarray}
Here $\alpha_1,\alpha_2,\alpha_3,\alpha_4$ are the possible permutations of of $t_1,t_2,0, t$, and 
\be
F_{t t_i}(\alpha_1,\alpha_2,\alpha_3,\alpha_4)=\langle e^{i \phi(\alpha_1) s_i(\alpha_1)} e^{i \phi(\alpha_2) s_i(\alpha_2)}e^{i \phi(\alpha_3) s_i(\alpha_3)}e^{i \phi(\alpha_4) s_i(\alpha_4)}\rangle ,
\ee
with $t_0=0$, and $s_0(0,t,t_1,t_2)=(1,1,-1,-1)$, $s_1(0,t,t_1,t_2)=(-1,1,1,-1)$, and $s_2(0,t,t_1,t_2)=(-1,1,-1,1)$.
More explicitly $F_{t t_i}$ is the four point correlation function of the four $e^{i \phi}$ operators in which $\phi(t)$ and $\phi(t_i)$ appear with the same sign in the exponent.

Using Wick's theorem we find that
\begin{eqnarray}
&F_{t t_i}&(\alpha_1,\alpha_2,\alpha_3,\alpha_4)=|\alpha_1-\alpha_2|^{\delta_{\phi} s_i(\alpha_1) s_i(\alpha_2)}  |\alpha_1-\alpha_3|^{\delta_{\phi} s_i(\alpha_1) s_i(\alpha_3)}|\alpha_1-\alpha_4|^{\delta_{\phi} s_i(\alpha_1) s_i(\alpha_4)}\times
\nonumber \\&
\times& |\alpha_2-\alpha_3|^{\delta_{\phi} s_i(\alpha_2) s_i(\alpha_3)}|\alpha_2-\alpha_4|^{\delta_{\phi} s_i(\alpha_2) s_i(\alpha_4)}|\alpha_3-\alpha_4|^{\delta_{\phi} s_i(\alpha_3) s_i(\alpha_4)}\times
\nonumber \\ &\times &\exp\{i \delta_{\phi} \pi [ sgn(\alpha_1-\alpha_2) s_i(\alpha_1) s_i(\alpha_2)+sgn(\alpha_1-\alpha_3) s_i(\alpha_1) s_i(\alpha_3)+sgn(\alpha_1-\alpha_4) s_i(\alpha_1) s_i(\alpha_4)+\nonumber \\&+& sgn(\alpha_2-\alpha_3) s_i(\alpha_2) s_i(\alpha_3)+sgn(\alpha_2-\alpha_4) s_i(\alpha_2) s_i(\alpha_4)+sgn(\alpha_3-\alpha_4) s_i(\alpha_3) s_i(\alpha_4)]/2\},
\end{eqnarray}
where $\delta_\phi=g/2$.

As detailed in Appendix B we also find the four point correlation function for the $\sigma$ operators starting from a CFT analysis of the correlation functions of the Ising model \cite{ginsparg}.
  \ba
&&\langle \sigma(\alpha_1)\sigma(\alpha_2) \sigma(\alpha_3) \sigma(\alpha_4) \rangle =F^{\sigma}_{12}(\alpha_1,\alpha_2,\alpha_3,\alpha_4)[\theta(1324)+\theta(1342)+\theta(2413)
 +\theta(2431)
+\theta(3124)+\theta(3142)
\nonumber \\&&  
+\theta(4213)+\theta(4231)]
\nonumber \\&&
+F^{\sigma}_{13}(\alpha_1,\alpha_2,\alpha_3,\alpha_4)[\theta(1234)+\theta(1432)+\theta(2143)
+\theta(2341)+\theta(3214)+\theta(3412)+\theta(4123)+\theta(4321)]
\nonumber \\&&
+F^{\sigma}_{14}(\alpha_1,\alpha_2,\alpha_3,\alpha_4)[\theta(1243)+\theta(1423)+\theta(2134)
+\theta(2314)+\theta(3241)+\theta(3421)+\theta(4132)+\theta(4312)],
\ea
where $\theta(abcd)=1$ for $\alpha_a>\alpha_b>\alpha_c>\alpha_d$ and is zero otherwise.
Also 
\ba
F^{\sigma}_{12}(\alpha_1,\alpha_2,\alpha_3,\alpha_4)&=&|\alpha_1-\alpha_2|^{\delta_{\sigma}}  |\alpha_3-\alpha_4|^{\delta_{\sigma}}|\alpha_1-\alpha_3|^{-\delta_{\sigma}} |\alpha_1-\alpha_4|^{-\delta_{\sigma}} |\alpha_2-\alpha_3|^{-\delta_{\sigma}}|\alpha_2-\alpha_4|^{-\delta_{\sigma} } \times \nonumber \\ &\times &e^{i \pi \delta_{\sigma}[sgn(\alpha_1-\alpha_2)+sgn(\alpha_3-\alpha_4)-
sgn(\alpha_1-\alpha_3)-sgn(\alpha_1-\alpha_4)-sgn(\alpha_2-\alpha_3)-sgn(\alpha_2-\alpha_4)]/2},
\ea
and similarly 
\ba
F^{\sigma}_{13}(\alpha_1,\alpha_2,\alpha_3,\alpha_4)&=&|\alpha_1-\alpha_3|^{\delta_{\sigma}}  |\alpha_2-\alpha_4|^{\delta_{\sigma}}|\alpha_1-\alpha_2|^{-\delta_{\sigma}} |\alpha_1-\alpha_4|^{-\delta_{\sigma}} |\alpha_2-\alpha_3|^{-\delta_{\sigma}}|\alpha_3-\alpha_4|^{-\delta_{\sigma} } \times \nonumber \\ &\times &e^{i \pi \delta_{\sigma}[sgn(\alpha_1-\alpha_3)+sgn(\alpha_2-\alpha_4)-
sgn(\alpha_1-\alpha_2)-sgn(\alpha_1-\alpha_4)-sgn(\alpha_2-\alpha_3)-sgn(\alpha_3-\alpha_4)]/2}, \ea          
and
\ba
F^{\sigma}_{14}(\alpha_1,\alpha_2,\alpha_3,\alpha_4)&=&|\alpha_1-\alpha_4|^{\delta_{\sigma}}  |\alpha_2-\alpha_3|^{\delta_{\sigma}}|\alpha_1-\alpha_3|^{-\delta_{\sigma}} |\alpha_1-\alpha_2|^{-\delta_{\sigma}} |\alpha_3-\alpha_4|^{-\delta_{\sigma}}|\alpha_2-\alpha_4|^{-\delta_{\sigma} } \times \nonumber \\ &\times &e^{i \pi \delta_{\sigma}[sgn(\alpha_1-\alpha_4)+sgn(\alpha_2-\alpha_3)-
sgn(\alpha_1-\alpha_3)-sgn(\alpha_1-\alpha_2)-sgn(\alpha_3-\alpha_4)-sgn(\alpha_2-\alpha_4)]/2},
\ea      
 with $\delta_{\sigma}=1/4$.
   
After some tedious but straightforward algebra one gets:   
\begin{eqnarray}
\langle S(t) \rangle_1=-2 {e^*}^2 |\Gamma|^4\Big( &\int_{t_1,t_2}^{t_1<0<t_2<t} &
(f_{tt_1}^{\delta_\phi}-f_{t0}^{\delta_\phi})\{\tilde{f}_{tt_2}^{\delta_\sigma}\cos \pi 
(\delta_\phi-\delta_\sigma)
+\tilde{f}_{t0}^{\delta_\sigma} [\cos \pi   
(\delta_\phi+\delta_\sigma)-2]\}
\nonumber
\\&&
+f_{tt_2}^{\delta_\phi}\{\tilde{f}_{tt_2}^{\delta_\sigma}\cos \pi 
(\delta_\phi+\delta_\sigma)
+\tilde{f}_{t0}^{\delta_\sigma} [\cos \pi   
(\delta_\phi+\delta_\sigma)-1-\cos 2 \pi \delta_\phi]\}
\nonumber \\
+&\int_{t_1,t_2}^{t_1<t_2<0<t}&
f_{tt_1}^{\delta_\phi} [\tilde{f}_{tt_2}^{\delta_\sigma}\cos \pi 
(\delta_\phi+\delta_\sigma)
-\tilde{f}_{t0}^{\delta_\sigma} \cos \pi   
(\delta_\phi-\delta_\sigma)]
\nonumber
\\&&
+f_{tt_2}^{\delta_\phi}[\tilde{f}_{tt_2}^{\delta_\sigma}\cos \pi 
(\delta_\phi+\delta_\sigma)
-\tilde{f}_{t0}^{\delta_\sigma} \cos \pi   
(\delta_\phi+\delta_\sigma)]
\nonumber
\\&&
+f_{t0}^{\delta_\phi}\{\tilde{f}_{t0}^{\delta_\sigma}\cos \pi 
(\delta_\phi+\delta_\sigma)
+\tilde{f}_{t2}^{\delta_\sigma} [1-\cos \pi   
(\delta_\phi+\delta_\sigma)-\cos 2 \pi \delta_\phi]\}
\nonumber \\
+&\int_{t_1,t_2}^{0<t_1<t_2<t}&
(f_{tt_1}^{\delta_\phi}-f_{t0}^{\delta_\phi})[\tilde{f}_{tt_1}^{\delta_\sigma}\cos \pi 
(\delta_\phi+\delta_\sigma)
-\tilde{f}_{tt_2}^{\delta_\sigma} \cos \pi   
(\delta_\phi-\delta_\sigma)]
\nonumber
\\&&
-f_{tt_2}^{\delta_\phi}\{\tilde{f}_{tt_2}^{\delta_\sigma}\cos \pi 
(\delta_\phi+\delta_\sigma)
+\tilde{f}_{tt_1}^{\delta_\sigma} [\cos 2 \pi \delta_\phi-1-\cos \pi   
(\delta_\phi+\delta_\sigma)]\}\Big)
\label{res}
\end{eqnarray}
where
\ba
 f_{t 0}^{\mu}&=&|t|^\mu |t_1-t_2|^{\mu} |t-t_1|^{-\mu} |t-t_2|^{-\mu}
 |t_1|^{-\mu} |t_2|^{-\mu} \cos[\omega_0(t-t_1-t_2)]
\nonumber \\
f_{t t_1}^{\mu}&=&|t-t_1|^\mu|t_2|^{\mu} |t|^{-\mu} |t-t_2|^{-\mu}
|t_1-t_2|^{-\mu} |t_1|^{-\mu}   \cos[\omega_0(t+t_1-t_2)]
\nonumber \\
f_{t t_2}^{\mu}&=&|t-t_2|^\mu |t_1|^{\mu} |t-t_1|^{-\mu} |t|^{-\mu}
|t_1-t_2|^{-\mu} |t_2|^{-\mu}   \cos [\omega_0(t+t_2-t_1)], 
\ea    
 $\tilde{f}^{\mu}_{t 0, t_1,t_2}=f^{\mu}_{t 0 t_1 t_2}(\omega_0\rightarrow 0)$, and for the Pfaffian state $\delta_{\sigma}=\delta_{\phi}=1/4$.

For comparison, we also provide the similar result for the case of a Laughlin state with filling factor $\nu$, obtained from the above equation by setting $\delta_\phi \rightarrow 2 \nu$ and $\delta_\sigma \rightarrow 0$:
\ba
\langle S(t) \rangle_1= &&-2 {e^*}^2 |\Gamma|^4 \{ \int_{t_1,t_2}^{t1_<0<t_2<t} [2 f_{t 0}^{\delta_l}(1-\cos \pi \delta_l)+2 f_{t t_1}^{\delta_l}(\cos \pi \delta_l-1)+f_{t t_2}^{\delta_l}(2 \cos \pi \delta_l-1-\cos 2 \pi \delta_l)] \nonumber \\&& +\int_{t_1,t_2}^{t1_<t_2<0<t} f_{t0}^{\delta_l}(1-\cos 2 \pi \delta_l) 
+\int_{t_1,t_2}^{0<t1_<t_2<t} f_{t t_2}^{\delta_l}(1-\cos 2 \pi \delta_l) \},
\ea
where $\delta_l=2 \nu$.

\section{Four point correlation function for the $\sigma$ operators}

In order to derive the real time four point correlation function for the Ising 
$\sigma$ operators, we start from the imaginary time conformal field theory result \cite{ginsparg}:
\be
\langle \sigma(z_1)  \sigma(z_2)\sigma(z_3)\sigma(z_4)\rangle= \Big| \frac{z_{13} z_{24}}{z_{14} z_{23} z_{12} z_{34}} \Big|^{1/4} \sqrt{\frac{1+|{\cal X}|+|1-{\cal X}|}{2}}, \ee
where $z=\tau+i x$, and $z_{i j}=z_i-z_j$. We chose to denote ${\cal X}=z_{12} z_{34}/z_{13} z_{24}$.
Since all our correlations will involve operators at the tunnelling point we will set all  $x_i$'s to zero.
In this limit we can evaluate the four point correlation functions and we note that
\ba 
\langle \sigma(z_1)  \sigma(z_2)\sigma(z_3)\sigma(z_4)\rangle=&&  \Big| \frac{z_{13} z_{24}}{z_{14} z_{23} z_{12} z_{34}} \Big|^{1/4},\text{ for } 0<{\cal X}<1,
\nonumber \\&&
 \Big| \frac{z_{13} z_{24}}{z_{14} z_{23} z_{12} z_{34}} \Big|^{1/4} \sqrt{1-{\cal X}}= \Big| \frac{z_{14} z_{23}}{z_{13} z_{24} z_{12} z_{34}} \Big|^{1/4}, \text{ for }{\cal X}<0,
\nonumber \\&&
 \Big| \frac{z_{13} z_{24}}{z_{14} z_{23} z_{12} z_{34}} \Big|^{1/4} \sqrt{\cal X}= \Big| \frac{z_{12} z_{34}}{z_{13} z_{24} z_{23} z_{14}} \Big|^{1/4}, \text{ for }{\cal X}>1,
\ea
By analytical continuation $\tau_i \rightarrow i \alpha_i$ to real time this yields:
\ba
&&\langle \sigma(\alpha_1)\sigma(\alpha_2) \sigma(\alpha_3) \sigma(\alpha_4) \rangle =F^{\sigma}_{12}(\alpha_1,\alpha_2,\alpha_3,\alpha_4)[\theta(1324)+\theta(1342)+\theta(2413)
 +\theta(2431)
+\theta(3124)+\theta(3142)
\nonumber \\&&  
+\theta(4213)+\theta(4231)]
\nonumber \\&&
+F^{\sigma}_{13}(\alpha_1,\alpha_2,\alpha_3,\alpha_4)[\theta(1234)+\theta(1432)+\theta(2143)
+\theta(2341)+\theta(3214)+\theta(3412)+\theta(4123)+\theta(4321)]
\nonumber \\&&
+F^{\sigma}_{14}(\alpha_1,\alpha_2,\alpha_3,\alpha_4)[\theta(1243)+\theta(1423)+\theta(2134)
+\theta(2314)+\theta(3241)+\theta(3421)+\theta(4132)+\theta(4312)],
\ea
where $\theta(abcd)=1$ for $\alpha_a>\alpha_b>\alpha_c>\alpha_d$ and is zero otherwise.
Also 
\ba
F^{\sigma}_{12}(\alpha_1,\alpha_2,\alpha_3,\alpha_4)&=&|\alpha_1-\alpha_2|^{\delta_{\sigma}}  |\alpha_3-\alpha_4|^{\delta_{\sigma}}|\alpha_1-\alpha_3|^{-\delta_{\sigma}} |\alpha_1-\alpha_4|^{-\delta_{\sigma}} |\alpha_2-\alpha_3|^{-\delta_{\sigma}}|\alpha_2-\alpha_4|^{-\delta_{\sigma} } \times \nonumber \\ &\times &e^{i \pi \delta_{\sigma}[sgn(\alpha_1-\alpha_2)+sgn(\alpha_3-\alpha_4)-
sgn(\alpha_1-\alpha_3)-sgn(\alpha_1-\alpha_4)-sgn(\alpha_2-\alpha_3)-sgn(\alpha_2-\alpha_4)]/2},
\ea
and similarly 
\ba
F^{\sigma}_{13}(\alpha_1,\alpha_2,\alpha_3,\alpha_4)&=&|\alpha_1-\alpha_3|^{\delta_{\sigma}}  |\alpha_2-\alpha_4|^{\delta_{\sigma}}|\alpha_1-\alpha_2|^{-\delta_{\sigma}} |\alpha_1-\alpha_4|^{-\delta_{\sigma}} |\alpha_2-\alpha_3|^{-\delta_{\sigma}}|\alpha_3-\alpha_4|^{-\delta_{\sigma} } \times \nonumber \\ &\times &e^{i \pi \delta_{\sigma}[sgn(\alpha_1-\alpha_3)+sgn(\alpha_2-\alpha_4)-
sgn(\alpha_1-\alpha_2)-sgn(\alpha_1-\alpha_4)-sgn(\alpha_2-\alpha_3)-sgn(\alpha_3-\alpha_4)]/2}, \ea          
and
\ba
F^{\sigma}_{14}(\alpha_1,\alpha_2,\alpha_3,\alpha_4)&=&|\alpha_1-\alpha_4|^{\delta_{\sigma}}  |\alpha_2-\alpha_3|^{\delta_{\sigma}}|\alpha_1-\alpha_3|^{-\delta_{\sigma}} |\alpha_1-\alpha_2|^{-\delta_{\sigma}} |\alpha_3-\alpha_4|^{-\delta_{\sigma}}|\alpha_2-\alpha_4|^{-\delta_{\sigma} } \times \nonumber \\ &\times &e^{i \pi \delta_{\sigma}[sgn(\alpha_1-\alpha_4)+sgn(\alpha_2-\alpha_3)-
sgn(\alpha_1-\alpha_3)-sgn(\alpha_1-\alpha_2)-sgn(\alpha_3-\alpha_4)-sgn(\alpha_2-\alpha_4)]/2},
\ea      
 with $\delta_{\sigma}=1/4$.

We can try to understand the physics of this result; one notes that $F_{12/13/14}^{\sigma}$ are the three cross ratios appearing in the fourth order correlation functions.
In a fourth order correlation function for an abelian state, they usually appear in quasi-symmetric combinations, 
such that by exchanging two of the times the most significant effect is the apparition of various phase factors, characteristic of fractional statistics. 
In the non-abelian case however, we see that the physics is very different, in that when we exchange two of the times, not only one acquires  phase factors, but one can also change the form of the correlation function from one of $F_{12/13/14}^{\sigma}$ cross ratios to another.

\end{widetext}

\end{document}